\documentclass[pra,aps,10pt,twocolumn,amsmath,amssymb]{revtex4-1}
\usepackage[english]{babel}
\usepackage{graphicx}
\usepackage{bm}
\usepackage{color}
\usepackage{slashed}
\usepackage{latexsym}
\usepackage{appendix}

\newcommand{\dd}{d}
\newcommand{\ii}{i}
\newcommand{\ee}{e}

\newcommand{\me}{m_\mathrm{e}} 
\newcommand{\op}[1]{\hat{#1}}
\newcommand{\ket}[1]{|{#1}\rangle}

\begin{document}

\title{Multiphoton ionization distributions beyond the dipole approximation:\\ Retardation versus recoil corrections}
\author{J. Z. Kami\'nski} \email{Jerzy.Kaminski@fuw.edu.pl}
\author{K. Krajewska} \email{Katarzyna.Krajewska@fuw.edu.pl}

\affiliation{
Institute of Theoretical Physics, Faculty of Physics, University of Warsaw, Pasteura 5, 02-093 Warsaw, Poland }
\date{\today}

\begin{abstract}
We study nondipole effects in multiphoton ionization of a two-dimensional hydrogen-like atom by a flat-top laser pulse of varied intensity.
For this purpose, we solve numerically a two-dimensional Schr\"odinger equation treating a propagating laser pulse exactly. The resulting
distributions are then compared to those calculated in the dipole approximation. A directional dependence of the energy-angular photoelectron 
distributions is demonstrated numerically in the case of a propagating laser pulse of a moderate and a high intensity. It is analytically interpreted
based on the leading order relativistic expansion of the electron Volkov state, showing a significant contribution of the electron recoil
to that behavior. In contrast, the retardation correction originating from the space- and time-dependence of the laser
field leads to a tiny redshift of the photoelectron energy spectra. Other features of ionization distributions are also analyzed,
including the sidelobes and the double-hump structures of multiphoton peaks, or their disappearance for intense propagating laser pulses.
\end{abstract}

\maketitle

\section{Introduction}
\label{sec:introduction}

The technology of producing coherent laser pulses has gone through various stages of development~\cite{RevModPhys.78.309}. Initially, these were 
light pulses with durations significantly exceeding the timescales typical for atomic processes. Additionally, the maximum electric field strengths 
were much smaller than the corresponding strengths met in atomic physics. Therefore, an approximate description of such pulses as monochromatic 
plane waves with random phases was justified. Consequently, the ionization was perceived as either a tunneling~\cite{Keldysh1964} 
or a multiphoton process~\cite{FHMFaisal_1973,PhysRevLett.42.1127,PhysRevA.22.1786}. In the latter case, the energy spectrum of photoelectrons 
generated in ionization is characterized by distinct peaks separated from each other by the frequency of the light wave, multiplied by the reduced 
Planck constant $\hbar$. This situation changed, however, when the laser technology was developed further to shorten the duration of light pulses to  
several electric field oscillations and to significantly increase their intensities. In such situations, not only the multiphoton picture of ionization
breaks down, but also nondipole effects related to light pressure start to appear. Such modifications of the multiphoton picture are inherently 
associated with an increasingly significant influence of the magnetic component of the laser field on the ongoing quantum process. 
For example, the angular distributions of photoelectrons are no longer centered around the laser field polarization vector, but also depend 
on the direction of pulse propagation, its intensity, central frequency, and the envelope phase. In particular, in the case of pulses with frequencies 
higher than the ionization potential and with energies of the electron quiver motion larger than the nonrelativistic atomic 
unit (but still lower than the electron rest mass energy, $\me c^2$), theoretical studies based on three- and two-dimensional spatial models indicate 
the formation of jet structures in the photoelectron momentum distributions, at least in their low-energy 
part~\cite{PhysRevLett.97.043601,PhysRevA.104.L021102,PhysRevA.107.053112}. It turns out that such structures significantly depend on the envelope 
phase and are generated as a result of the electron wave packet spreading after the interaction with the laser pulse. This process is modified by 
the post-pulse interaction of photoelectrons with their parent ions~\cite{PhysRevA.107.053112} (as opposed to the in-pulse rescattering phenomenon 
observed in the dipole approximation~\cite{Potvliege2011,Milošević_2006,Becker_2018}, when the transfer of momentum from the laser field is not accounted for) and leads to rich interference structures resulting from the generation of the so-called quantum vortex streets \cite{Suster:24} (similar to the von K\'arm\'an vortex streets observed in hydrodynamic flows, but also predicted in Bose-Einstein condensates \cite{PhysRevLett.104.150404}, ionization \cite{Larionov2018} or photodetachment by linearly \cite{PhysRevA.102.043102} and circularly \cite{PhysRevA.104.033111} polarized pulses).

The aim of our work is to investigate conditions under which the ionization picture, based on the multiphoton absorption, collapses in the case of 
long laser pulses with a central frequency larger than the ionization potential of atoms. We analyze this problem by investigating the time evolution 
of a single-active-electron system described by the Schr\"odinger equation, in which the laser field is assumed to be a propagating in space 
electromagnetic pulse with a flat front. Our goal is to study numerically this process in situations in which the leakage of the electron wave 
function from the integration region is negligibly small after the interaction with the laser pulse. We insist upon studying cases, in which in 
the dipole approximation we observe the absorption of over 100 photons and the energy spectrum of photoelectrons also accounts for the effects 
related to the rescattering of electrons in the pulse (see, e.g., \cite{Milošević_2006} and references therein). For this reason, we limit our 
numerical analysis to two-dimensional systems. However, the physical interpretation of our results is based on the strong-field approximation (SFA), that is
valid for arbitrarily dimensional spaces in the high-energy portion of the photoelectron energy spectrum, i.e., when the Born approximation for scattering states is applicable.

This paper is organized as follows. In Sec.~\ref{sec:theory} we present the general framework of our theoretical investigations. Our numerical method 
used for the time-evolution of initially bound electrons is described in Sec.~\ref{sec::method}. In Sec.~\ref{sec::model}, we specify the binding
potential and the vector potential of the laser pulse. Some details regarding parameters used in our numerical studies are provided in 
Sec.~\ref{sec:numeric}, whereas the definition of the high-energy photoelectron distributions are given in Sec.~\ref{sec:spectrum}. 
In Sec.~\ref{sec:ModeratePulse}, we consider the ionization process by moderately intense pulses and compare the results with the ones that are calculated 
within the dipole approximation. The interpretation of our findings based on the SFA is discussed in Sec.~\ref{sec:SFAanalysis}, in which 
we also explain quantitatively what the terms 'weak', 'moderate', and 'strong' laser pulses mean. The analysis for strong laser pulses is presented 
in Sec.~\ref{sec:StrongPulse}, where we show the collapse of the concept of multiphoton peaks in the photoelectron energy spectra, even though 
it still remains useful in the dipole approximation. Finally, in Sec.~\ref{sec:conclusion} we draw our conclusions.

In our numerical analysis, we use the atomic units of momentum $p_0=\alpha\me c$, energy $E_0=\alpha^2\me c^2$, length $a_0=\hbar/p_0$, time $t_0=\hbar/E_0$, and the electric field strength $\mathcal{E}_0=\alpha^3\me^2 c^3/(|e|\hbar)$, where $\me$ and $e=-|e|$ are the electron rest mass and charge, and $\alpha$ is the fine-structure constant. In analytical formulas we put $\hbar=1$, while keeping explicitly the remaining fundamental constants.

\section{Theoretical framework}
\label{sec:theory}

We consider a two-dimensional hydrogen-like atom exposed to a linearly polarized laser pulse. Within a single-active-electron
approximation, it is described by the minimal coupling Hamiltonian,
\begin{equation}
\op{H}(\op{\bm{p}},\op{\bm{x}},t)=\frac{1}{2\me}\bigl(\op{\bm{p}}-e\bm{A}(\op{\bm{x}},t)\bigr)^2+V(\op{\bm{x}}),
\label{ps4}
\end{equation}
where $\op{\bm x}$ and $\op{\bm p}$ are the position and momentum operators, respectively, $V(\op{\bm{x}})$ represents the potential energy of electron-ion interaction, whereas
\begin{equation}
\bm{A}(\op{\bm{x}},t)=(A(t-\op{x}_2/c),0)
\label{ps5}
\end{equation}
is the vector potential of the laser pulse. Eq.~\eqref{ps5} represents the pulse propagating along the $x_2$-direction, that is linearly polarized 
in the $x_1$-direction. As explained in Ref.~\cite{PhysRevA.107.053112}, in such case it is possible to solve numerically the time-dependent Schr\"odinger equation (TDSE),
\begin{equation}
\ii\partial_t\ket{\psi(t)}=\op{H}(\op{\bm{p}},\op{\bm{x}},t)\ket{\psi(t)},
\label{th1}
\end{equation}
by adapting the Suzuki-Trotter method. For convenience of the reader, we briefly reiterate the method now. More details can be found in Ref.~\cite{PhysRevA.107.053112}. 

\subsection{Numerical method}
\label{sec::method}

The solution of the Schr\"odinger equation~\eqref{th1} can be written as
\begin{equation}
\ket{\psi(t)}=\op{U}(t,t_0)\ket{\psi_0(t_0)},\label{sol}
\end{equation}
where $\ket{\psi_0(t_0)}$ represents the initial state of the system, whereas
$\op{U}(t,t_0)$ is the corresponding time evolution operator. It satisfies the same wave equation,
\begin{equation}
\ii\partial_t\op{U}(t,t_0)=\op{H}(\op{\bm{p}},\op{\bm{x}},t)\op{U}(t,t_0),
\label{th4}
\end{equation}
which can be formally solved,
\begin{equation}
\op{U}(t,t_0)=\op{\mathcal{T}}\Bigl[\exp\Bigl(-\ii\int_{t_0}^t \dd\tau \op{H}(\op{\bm{p}},\op{\bm{x}},\tau)\Bigr)\Bigr].
\label{th5}
\end{equation}
Here, $\op{\mathcal{T}}$ is the Dyson time-ordering operator. In practice, we introduce time discretization,
\begin{equation}
t_n=t_0+n\delta t, \quad n=0,1,\dots ,N, \quad \delta t=\frac{t-t_0}{N},
\label{th7}
\end{equation}
which allows us to represent the time evolution operator as 
\begin{equation}
\op{U}(t,t_0)=\prod_{n=0}^{N-1}\op{U}(t_{n+1},t_n).
\label{th8}
\end{equation}
For a sufficiently small time increment $\delta t$, we can use the approximation,
\begin{equation}
\op{U}(t_{n+1},t_{n})=\ee^{-\ii\delta t \op{H}(\op{\bm{p}},\op{\bm{x}},\bar{t}_n)},
\label{th9}
\end{equation}
with $\bar{t}_n=(t_{n+1}+t_n)/2$. To calculate this exponent, we split the Hamiltonian~\eqref{ps4},
\begin{equation}
\op{H}(\op{\bm{p}},\op{\bm{x}},t)=\op{H}_1(\op{\bm{p}},t)+\op{H}_2(\op{\bm{x}},t)+\op{h}_3(\op{\bm{p}},\op{\bm{x}},t),
\label{ps6}
\end{equation}
such that
\begin{align}
\op{H}_1(\op{\bm{p}},t)=&\frac{1}{2\me}\op{\bm{p}}^2,\label{ps77} \\ 
\op{H}_2(\op{\bm{x}},t)=&\frac{1}{2\me}\bigl(eA(t-\op{x}_2/c)\bigr)^2+V(\op{\bm{x}}), \label{ps7}\\
\op{h}_3(\op{\bm{p}},\op{\bm{x}},t)=&-\frac{e}{\me}A(t-\op{x}_2/c)\cdot\op{p}_1. \label{ps8}
\end{align}
Using the Suzuki-Trotter formula, we find out that an infinitesimal evolution operator $\op{U}(t_{n+1},t_n)$ takes the form
\begin{equation}
\op{U}(t_{n+1},t_n)=\op{U}_{{\rm ST}3}(t_{n+1},t_n)+O((\delta t)^3),
\label{th14}
\end{equation}
where
\begin{align}
\op{U}_{{\rm ST}3}(t_{n+1},t_n)&=\ee^{-\ii\frac{\delta t}{2} \op{H}_1(\op{\bm{p}},\bar{t}_n)}\ee^{-\ii\frac{\delta t}{2} \op{H}_2(\op{\bm{x}},\bar{t}_n)} 
 \ee^{-\ii\delta t \op{h}_3(\op{\bm{p}},\op{\bm{x}},\bar{t}_n)}\nonumber\\
 &\times\ee^{-\ii\frac{\delta t}{2}\op{H}_2(\op{\bm{x}},\bar{t}_n)}\ee^{-\ii\frac{\delta t}{2}\op{H}_1(\op{\bm{p}},\bar{t}_n)}.
\label{th15}
\end{align}
Finally, given the decomposition~\eqref{ps6}, the time evolution operator~\eqref{th5} can be calculated as
\begin{equation}
\op{U}(t,t_0)=\prod_{n=0}^{N-1}\op{U}_{{\rm ST}3}(t_{n+1},t_n)+O((\delta t)^2),
\label{th16}
\end{equation}
with an overall numerical error of the order of $(\delta t)^2$.

It is important to note that the above scheme allows for efficient numerical calculations. This follows from the fact that each exponent in
Eq.~\eqref{th15} can be efficiently applied to any state of the system. This concerns also the exponent with the mixed (i.e., position-momentum) 
representation of the Hamiltonian $\op{h}_3(\op{\bm p},\op{\bm x},t)$ [Eq.~\eqref{ps8}]. As elaborated in detail in Ref.~\cite{PhysRevA.107.053112}, for the chosen
Hamiltonian decomposition~\eqref{ps6}, the split-step Fourier method can be conveniently applied (see also Refs.~\cite{Liu:21,Hu:22}).

Note that in the dipole approximation, in which the vector potential \eqref{ps5} depends only on time, in principle the same numerical scheme can be applied. However, in this case, it is more efficient to use the $\op{U}_{{\rm ST}2}$ scheme described in \cite{PhysRevA.107.053112}.

\subsection{Theoretical model}
\label{sec::model}

In the following, we consider the electron-ion effective potential energy,
\begin{equation}
V(\bm{x})=-\mathcal{Z}(x)\alpha c\frac{\mathrm{erf}(x/a_V)}{x},
\label{ps1}
\end{equation}
where $\bm{x}=(x_1,x_2)$ is the two-dimensional position vector with the norm $x=|\bm{x}|=\sqrt{x_1^2+x_2^2}$, $\mathrm{erf}(z)$ is the error function, 
whereas the effective atomic number equals $\mathcal{Z}(x)=1-\lambda_V\exp(-x^2/b^2_V)$. The role of the error function is to soften the singularity 
at $x=0$ and to guarantee that for large $x$ the potential has the Coulomb tail $-1/x$. Moreover, $\mathcal{Z}(x)$ is chosen such that the energy 
of the ground state $E_B$ supported by $V({\bm x})$ is close to the ground state energy of the 3D hydrogen atom, $-0.5E_0$. This goal is achieved for 
$a_V=0.1a_0$, $b_V=10a_0$, and $\lambda_V=0.46$. The corresponding bound state wave function, $\psi_B(\bm{x})$, is calculated numerically using the Feynman-Kac method for imaginary times. In this way, we solve the eigenvalue problem for the atomic Hamiltonian,
\begin{equation}
\op{H}_{\rm at}(\op{\bm p},\op{\bm x})=\frac{1}{2\me}\op{\bm{p}}^2+V(\op{\bm{x}}),
\end{equation}
which provides the initial state for the time-propagation of the Schr\"odinger equation~\eqref{th1}. Note that similar regularization of the effective 
interaction has been used, for instance, in Refs.~\cite{shin1995nonadiabatic,erdmann2003combined,PhysRevA.102.013104}.

For the laser pulse introduced in Eq.~\eqref{ps5}, we choose the vector potential with a super Gaussian envelope, i.e.,
\begin{equation}
A(t)=A_0\exp\Bigl[-\Bigl(\eta\, \frac{\omega t-\pi N_\mathrm{osc}}{\pi N_\mathrm{osc}}\Bigr)^{N_\mathrm{env}}\Bigr]\sin(\omega t).
\label{ps14}
\end{equation}
In our numerical analysis, we consider a high-frequency pulse with the carrier frequency $\omega=2E_0$, whereas $\eta=1.35$ and $N_\mathrm{env}=12$ 
are chosen such that $A(t)$ vanishes for $t<0$ and $t>2\pi N_\mathrm{osc}/\omega$ within the accuracy of our numerical calculations. In this case, the envelope
remains nearly constant for $(N_\mathrm{osc}-2)$ cycles and so the laser pulse resembles a flat-top pulse. The remaining parameters, $A_0$ and 
$N_\mathrm{osc}$, will be defined below.

\subsection{Numerical details}
\label{sec:numeric}

Our numerical calculations are performed in a 2D spatial region determined by the parameters $x_{i0}$ ($i=1,2$) such that $-x_{i0}\leqslant x_i < x_{i0}$. 
The number of grid points in each direction is fixed by two integers $K_i$,
\begin{equation}
x_{i,j}=-x_{i0}+(j-1)\Delta x_i, \, \Delta x_i=\frac{2x_{i0}}{2^{K_i}}, \, j=1,\dots, 2^{K_i},
\label{nd1}
\end{equation}
for $i=1,2$. Consequently, the discretization in momentum space corresponds to
\begin{equation}
p_{i,j}=-\frac{\pi 2^{K_i}}{2x_{i0}}+(j-1)\Delta p_i,\, \Delta p_i=\frac{\pi}{x_{i0}}.
\label{nd2}
\end{equation}
The values for $K_i$ and $x_{i0}$ will be given below.

In order to reduce the boundary reflection effects we introduce the mask function $M(x_1,x_2)$, 
\begin{equation}
M(x_1,x_2)=\begin{cases}
1, & \tilde{x}<r_1, \cr
\bigl[\cos\bigl(\frac{\pi}{2}\frac{\tilde{x}-r_1}{r_2-r_1}\bigr)\bigr]^{1/8}, & r_1\leqslant \tilde{x}\leqslant r_2, \cr
0, & \tilde{x}>r_2,
\end{cases}
\label{nd3}
\end{equation}
where $\tilde{x}=x_0\sqrt{x_1^2/x_{10}^2+x_2^2/x_{20}^2}$ and $x_0=\mathrm{max}(x_{10},x_{20})$. This function multiplies 
$\exp(-\ii\frac{\delta t}{2} H_2(\bm{x},\bar{t}_n))$ in Eq.~\eqref{th15}. The remaining two parameters, $r_1$ and $r_2$, will be specified below.

As explained in Sec.~\ref{sec::method}, our numerical code for solving TDSE uses the Suzuki-Trotter scheme with the split-step Fourier approach to propagate in time the initial electron state. This is done starting at time $-x_{20}/c$, with time steps of $\delta t$. Moreover, we propagate for times not smaller than $2\pi N_{\rm osc}/\omega + x_{20} /c$. In other words, we make sure that the initial and final electron states are laser-field free. The convergence of our results is checked by varying parameters of our simulations. 

\subsection{Photoelectron momentum distribution}
\label{sec:spectrum}

In this paper, we focus on high-energy photoelectrons. In order to determine their momentum distributions we apply the method developed in 
Refs.~\cite{PhysRevA.92.051401,Ma:21}. Firstly, we propagate the initial ground state $\psi(\bm{x},t=-x_{20}/c)=\psi_B(\bm{x})$ to the final 
time $T_\mathrm{f}>2\pi N_{\rm osc}/\omega + x_{20} /c$. Then, the final electron wave packet, $\psi(\bm{x},T_\mathrm{f})$, is multiplied by 
a mask function, 
\begin{equation}
M_0(x)=\begin{cases}
0, & x<R_1, \cr
\frac{1}{2}\bigl[1-\cos\bigl(\pi\frac{x-R_1}{R_2-R_1}\bigr)\bigr], & R_1\leqslant x\leqslant R_2, \cr
1, & x>R_2,
\end{cases}
\label{nd4}
\end{equation}
with appropriately chosen values of $R_1$ and $R_2$. The role of $M_0(x)$ is to eliminate from the final electron wave packet the contributions 
coming from the bound and low-energy scattering states. In doing so, we obtain the truncated electron wave packet,
\begin{equation}\label{nd5}
\psi_m(\bm{x},T_\mathrm{f})=M_0(x)\psi(\bm{x},T_\mathrm{f}).
\end{equation}
Its Fourier transform, $\tilde{\psi}_m(\bm{p},T_\mathrm{f})$, defines the photoelectron momentum distribution,
\begin{equation}\label{nd6}
\tilde{P}_m(p_1,p_2,T_\mathrm{f})=|\tilde{\psi}_m(\bm{p},T_\mathrm{f})|^2.
\end{equation}
The latter is normalized such that
\begin{equation}\label{nd7}
\int\dd p_1\dd p_2 \tilde{P}_m(p_1,p_2,T_\mathrm{f})\leqslant 1,
\end{equation}
where the equality holds for $R_1=R_2=0$ provided that the leakage of the wave function from the integration region is negligibly small.

\begin{figure}
\includegraphics[width=7cm]{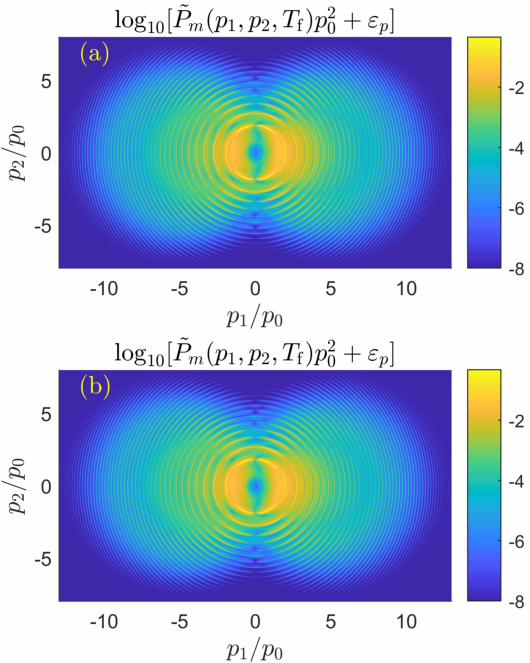}
\caption{Photoelectron momentum distributions plotted in the Cartesian coordinates for parameters defined in Sec.~\ref{sec:ModeratePulse}. 
In panel (a) we show the distribution calculated in the dipole approximation, whereas in panel (b) for the laser pulse propagating in 
the $x_2$-direction. Both distributions are shown in the logarithmic scale with $\varepsilon_p=10^{-8}$, meaning that we eliminate from the figures values 
smaller than $\varepsilon_p$. In both cases we observe multiphoton rings and the differences between panels (a) and (b) are hardly visible.
}
\label{MomentumDistr1Ca5xyK}
\end{figure}

In our analysis we also determine, by applying the cubic spline interpolation, the photoelectron distribution in the polar coordinates 
$(|\bm{p}|,\varphi_{\bm{p}})$, such that $p_1=|\bm{p}|\cos(\varphi_{\bm{p}})$ and $p_2=|\bm{p}|\sin(\varphi_{\bm{p}})$. However, in order 
not to complicate the notation, we will use the same symbols for these two distributions, i.e., $\tilde{P}_m(p_1,p_2,T_\mathrm{f})$ 
and $\tilde{P}_m(|\bm{p}|,\varphi_{\bm{p}},T_\mathrm{f})$, distinguishing them by explicitly writing their arguments, 
\begin{equation}\label{nd7a}
\tilde{P}_m(|\bm{p}|,\varphi_{\bm{p}},T_\mathrm{f})=\tilde{P}_m(p_1,p_2,T_\mathrm{f}).
\end{equation}
This also allows us to define the marginal distributions,
\begin{align}
\tilde{P}_m(|\bm{p}|,T_\mathrm{f})&=\int_0^{2\pi} \dd \varphi_{\bm{p}}\, \tilde{P}_m(|\bm{p}|,\varphi_{\bm{p}},T_\mathrm{f}),\label{nd8}\\
\tilde{P}_m(\varphi_{\bm{p}},T_\mathrm{f})&=\int_0^\infty |\bm{p}|\dd |\bm{p}|\, \tilde{P}_m(|\bm{p}|,\varphi_{\bm{p}},T_\mathrm{f}),\label{nd9}
\end{align}
and
\begin{equation}\label{nd10}
\tilde{P}_m(\varphi_{\bm{p}},E_{\mathrm{min}},T_\mathrm{f})=\int_{p_{\mathrm{min}}}^\infty |\bm{p}|\dd |\bm{p}|\, \tilde{P}_m(|\bm{p}|,\varphi_{\bm{p}},T_\mathrm{f}),
\end{equation}
where $E_{\mathrm{min}}=p_{\mathrm{min}}^2/2\me$. While Eq.~\eqref{nd8} defines the energy distribution, Eq.~\eqref{nd9} determines the angular
distribution, Eq.~\eqref{nd10} is the angular distribution of photoelectrons of kinetic energies $E_{\bm{p}}=\bm{p}^2/2\me$ larger then 
$E_{\mathrm{min}}$. Again, we use the same symbol for all these distributions and distinguish them by their arguments.

\section{Moderately intense laser pulses}
\label{sec:ModeratePulse}

We start with a long laser pulse, comprising of $N_\mathrm{osc}=20$ cycles, including 18 cycles within the flat-top portion of the envelope. The pulse central 
frequency $\omega=2E_0$ and the amplitude are such that $|e|A_0=5p_0$. This means that, within the pulse flat top, the ponderomotive energy equals 
$U_p=e^2A_0^2/4\me=6.25E_0$ and is larger then $\omega$. The convergence of numerical analysis is achieved for the following parameters: 
$x_{10}=x_{20}=800a_0$, $K_1=K_2=13$, and $\delta t=0.01t_0$. For the mask function $M(x_1,x_2)$ we choose $r_1=750a_0$ and $r_2=799a_0$. 
For these parameters the pulse lasts for $T_\mathrm{p}=2\pi N_\mathrm{osc}/\omega=62.83t_0$. The final time of the numerical calculation is 
chosen to be $T_\mathrm{f}=74.95t_0$. In this case, the probability for the electron to escape the integration region is smaller than 
$7\times 10^{-8}$, whereas the probability that it stays in the ground state equals $0.2$. For the mask function $M_0(x)$ we take $R_1=30a_0$ and $R_2=90a_0$.

\begin{figure}
\includegraphics[width=7cm]{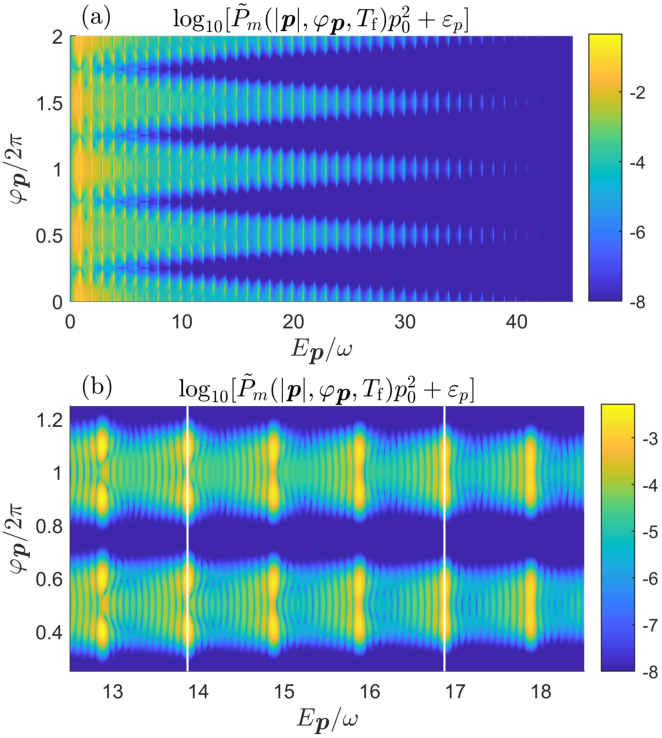}
\caption{The same as in Fig.~\ref{MomentumDistr1Ca5xyK}(a) but presented in the polar coordinates. For visual purposes, in panel (a) we show two copies of the distribution, i.e., 
$0\leqslant\varphi_{\bm{p}}\leqslant 4\pi$, whereas in panel (b) the enlarged portion of this distribution is demonstrated. We clearly observe vertical stripes separated by $\omega$ in the photoelectron 
energy scale ($E_{\bm{p}}=\bm{p}^2/2\me$), which correspond to multiphoton peaks. The meaning of the two vertical lines is discussed in Sec.~\ref{sec:SFAanalysis}.
}
\label{Interpolation800x800XY2RPhi1Cda5K}
\end{figure}

In Fig.~\ref{MomentumDistr1Ca5xyK} we present the photoelectron momentum distributions in the Cartesian coordinates calculated in the dipole 
approximation [panel (a)] and for the laser pulse propagating in the vertical $x_2$-direction [panel (b)]. In both panels we observe practically 
the same pattern, i.e., multiphoton rings of large probability values. This, however, does not mean that the nondipole effects are marginally small. The differences can be clearly seen if the distributions are presented in the polar coordinates.

\begin{figure}
\includegraphics[width=7cm]{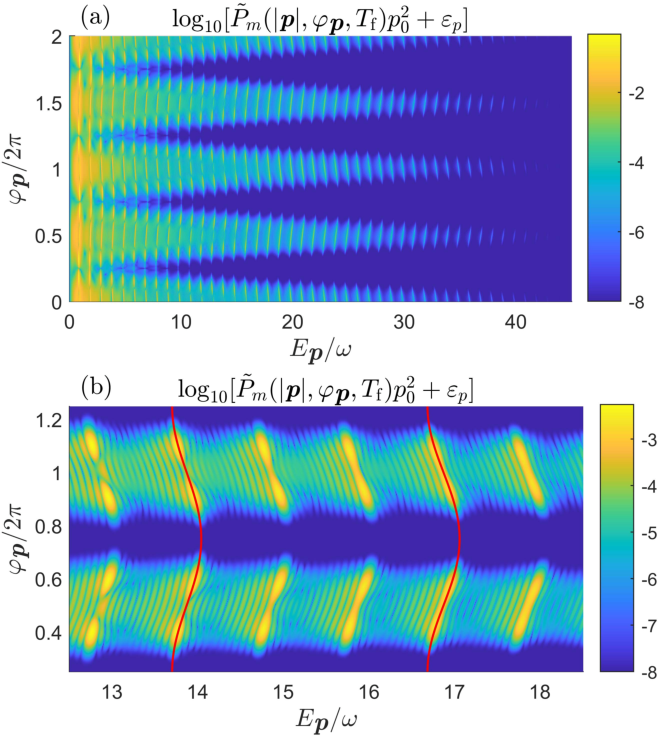}
\caption{The same as in Fig.~\ref{MomentumDistr1Ca5xyK}(b) but in the polar coordinates. For visual purposes, we show two copies of the distribution, 
i.e., $0\leqslant\varphi_{\bm{p}}\leqslant 4\pi$, in panel (a) and its enlarged portion in panel (b). Contrary to the results obtained within the dipole
approximation (Fig.~\ref{Interpolation800x800XY2RPhi1Cda5K}), we observe tilted stripes in the momentum distribution of emitted electrons. The meaning 
of the two wavy lines is discussed in Sec.~\ref{sec:SFAanalysis}.
}
\label{Interpolation800x800XY2RPhi1Ca5K}
\end{figure}

In Fig.~\ref{Interpolation800x800XY2RPhi1Cda5K} we represent the photoelectron momentum distribution shown in Fig.~\ref{MomentumDistr1Ca5xyK}(a), but 
this time in the polar coordinates. Those results have been calculated using the dipole approximation. We clearly see the vertical stripes separated by $\omega$;
thus, confirming the multiphoton ionization picture for the current pulse parameters. In between the main peaks we also observe much less intense 
secondary peaks, which are due to the interference originating from the finite character of the laser pulse. Verticality of these stripes means that, 
independently of the emission direction, a particular multiphoton peak appears for the same energy. Moreover, for energies $E_{\bm{p}}<20\omega$ 
(or $E_{\bm{p}}<6.4U_p$), the angular distribution of these peaks has a sidelobe structure. It is commonly understood that such structures are due 
to the in-pulse rescattering process. Namely, after separation from atoms, electrons return and instead of being captured along with the emission of
the high-order harmonics, they are scattered by their parent ions. As a net result, they escape to infinity in the direction that is not parallel 
to the laser field polarization axis. As shown in Fig.~\ref{Interpolation800x800XY2RPhi1Cda5K}(a) for larger energies, the sidelobes disappear and 
electrons are emitted in the polarization direction. This means that the rescattering process is much less probable for high-energy electrons.

The multiphoton picture of ionization is modified when the process is driven by a propagating laser pulse. The momentum distribution plotted in 
Fig.~\ref{MomentumDistr1Ca5xyK}(b) is now shown in the polar coordinates in Fig.~\ref{Interpolation800x800XY2RPhi1Ca5K}. Here, instead of vertical 
structures we observe the tilted stripes. However, as their inclination is small, we are still able to assign a multiphoton line (corresponding to the absorption 
of a certain number of photons of energy $\omega$) to the ionization process. This is the qualitative meaning of the term 'moderately intense laser 
pulses', used in the title of this section. The quantitative meaning of this term will be worked out based on the SFA approach in Sec.~\ref{sec:SFAanalysis}. 

\begin{figure}
\includegraphics[width=7cm]{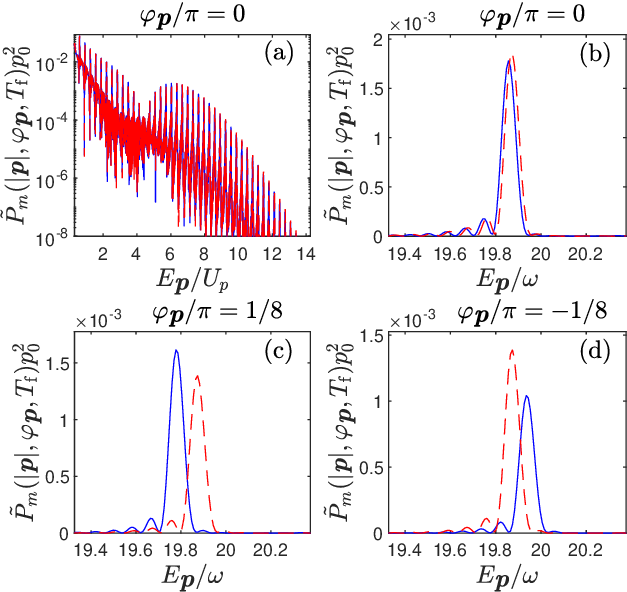}
\caption{Photoelectron momentum distributions in the polar coordinates, $\tilde{P}_m(|\bm{p}|,\varphi_{\bm{p}},T_\mathrm{f})$, for selected directions 
of electron emission, as indicated above each panel. The solid lines represent distributions for a propagating laser pulse, whereas the dashed 
lines correspond to the distributions calculated in the dipole approximation. For $\varphi_{\bm{p}}=0$ [panel (a)] a sort of plateau is observed, which is 
due to the in-pulse rescattering processes. Analysis of individual multiphoton peaks indicate both red- [for $\varphi_{\bm{p}}>0$, panel (c)] 
and blueshift [for $\varphi_{\bm{p}}<0$, panel (d)] due to the propagation of the driving pulse, with a tiny redshift for 
$\varphi_{\bm{p}}=0$ [panel (b)].
}
\label{Interpol800x800XY2RPhi1Ca5IntE}
\end{figure}

In Fig.~\ref{Interpol800x800XY2RPhi1Ca5IntE} we present the directional photoelectron momentum distributions, calculated for both 
the propagating laser pulse (solid lines) and within the dipole approximation (dashed lines). In Fig.~\ref{Interpol800x800XY2RPhi1Ca5IntE}(a), 
the distribution for the broad range of photoelectron energies is shown in the case when the electron is emitted along the polarization axis
of the pulse ($\varphi_{\bm{p}}=0$). We observe the nonmonotonic behavior of this distribution, which indicates the nonperturbative character 
of ionization for the current field parameters. The appearance of the plateau (or rather the monticule) centered around the value $E_{\bm{p}}=6.5U_p$ 
is due to the in-pulse rescattering. In the remaining panels, we investigate the particular multiphoton peak from that region.
For $\varphi_{\bm{p}}=0$, the peak under consideration almost coincides with its dipole
analogue [Fig.~\ref{Interpol800x800XY2RPhi1Ca5IntE}(b)]. We observe only a tiny redshift of the peak due to the nondipole effects. However, for the polar 
angles $\varphi_{\bm{p}}=\pi/8$ [Fig.~\ref{Interpol800x800XY2RPhi1Ca5IntE}(c)] and $\varphi_{\bm{p}}=-\pi/8$ [Fig.~\ref{Interpol800x800XY2RPhi1Ca5IntE}(d)], 
we detect correspondingly significant red- and blueshift of the peak once the propagation of the laser pulse is accounted for. The physical interpretation 
of these findings is given in Sec.~\ref{sec:SFAanalysis}.

\begin{figure}
\includegraphics[width=7cm]{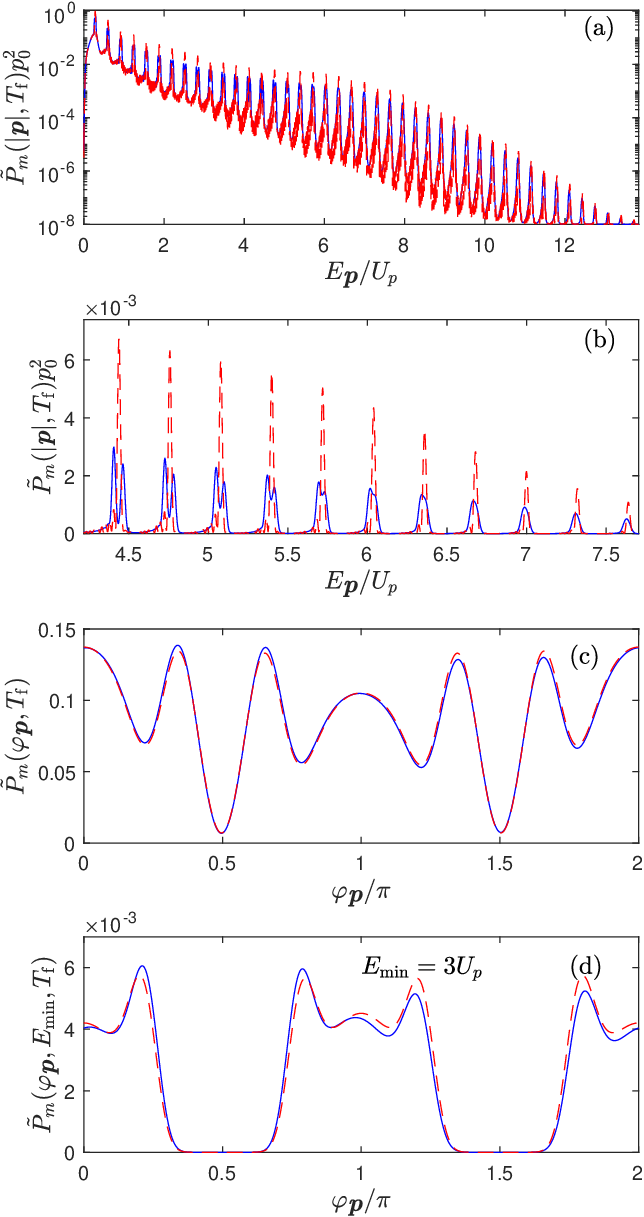}
\caption{Photoelectron marginal distributions plotted in the polar coordinates for parameters defined in Sec.~\ref{sec:ModeratePulse}. In panel (a) 
we show the energy distribution of photoelectrons, Eq.~\eqref{nd8}, with details 
presented in panel (b). We clearly see the multiphoton character of these distributions with well separated peaks, even though for the propagating laser 
pulse (solid lines) the nondipole effects significantly modify the peak structure. In panels (c) and (d) we demonstrate the angular distributions of 
photoelectrons defined by Eqs.~\eqref{nd9} and~\eqref{nd10}, respectively. We observe that in the direction of propagation of the laser pulse,
 i.e., for $0<\varphi_{\bm{p}}<\pi$, the spectra presented in panels (c) and (d) are slightly enhanced as compared to the dipole results, whereas they are  
 reduced in the opposite direction (i.e., for $\pi<\varphi_{\bm{p}}<2\pi$). The results in panel (d) have been calculated for $E_{\rm min}=3U_p$.
}
\label{Interpol800x800XY2RPhi1Ca5IntE0}
\end{figure}

In closing this section, let us discuss the marginal distributions presented in Fig.~\ref{Interpol800x800XY2RPhi1Ca5IntE0}, which have been calculated for both 
the propagating pulse (solid lines) and in the dipole approximation (dashed lines). In Fig.~\ref{Interpol800x800XY2RPhi1Ca5IntE0}(a), we demonstrate the 
polar-angle-integrated energy distribution of photoelectrons, Eq.~\eqref{nd8}. Here, we observe well separated multiphoton peaks together 
with the rescattering plateau, which ends up for electron energies around $E_{\bm{p}}=8U_p$. A closeup on the portion of this spectrum reveals additional 
substructure of the peaks in the case of a propagating laser pulse [solid line in Fig.~\ref{Interpol800x800XY2RPhi1Ca5IntE0}(b)]. More specifically, for energies smaller than 
$6U_p$ we observe double-hump peaks. This is a compound effect, which originates from the sidelobes observed in the photoelectron angular distributions in the dipole approximation
and the 'tilting of stripes' discussed above, that is due to nondipole corrections. The second effect occurs for larger electron energies and leads 
to a broadening of the peaks. However, on average, for all peaks we observe the redshift of the nondipole energy distribution as compared to the dipole one. 
Furthermore, the effects related to the radiation pressure, or to the momentum transfer from the propagating laser pulse to photoelectrons, are presented 
for the energy-integrated angular distribution~\eqref{nd9} in Fig.~\ref{Interpol800x800XY2RPhi1Ca5IntE0}(c). As we observe for the emission angles 
$0<\varphi_{\bm{p}}<\pi$, the nondipole corrections lead to a small enhancement of the ionization probability distribution. In contrast, for 
$\pi <\varphi_{\bm{p}}<2\pi$, we see the opposite behavior. Fig.~\ref{Interpol800x800XY2RPhi1Ca5IntE0}(d) shows that this trend is preserved also for the large 
energy portion of the distribution, defined by Eq.~\eqref{nd10} for $E_{\rm min}=3U_p$.

\section{SFA analysis}
\label{sec:SFAanalysis}

It is known that SFA, although simple in implementation in the lowest order, is nevertheless useful for interpretation of results obtained by a 
more sophisticated numerical analysis~\cite{ehlotzky1998electron,Popruzhenko2014Usp,milovsevic2024asymptotic}. The most spectacular example of such 
application is the interpretation of high-order harmonic generation~\cite{PhysRevA.49.2117}. It turns out that SFA can also be used in the analysis 
and interpretation of the results presented above. 

In zeroth order approximation, SFA consists in replacing in the expression for the ionization probability amplitude the exact scattering electron
state with incoming spherical waves by the Volkov solution~\cite{Wolkow1935}. In other words, it relies on applying the lowest-order Born approximation 
with respect to the static binding potential. If we want to include corrections resulting from the propagation of the laser pulse while still neglecting 
spin effects, this should be the Volkov solution of the Klein-Gordon equation \cite{Ehlotzky_2009},
\begin{equation}\label{sfa1}
\psi_{\bm{p}}(\bm{x},t)=\exp\bigl[-\ii p\cdot x+\ii\Phi_{\bm{p}}(\bm{x},t)\bigr],
\end{equation}
with
\begin{equation}\label{sfa2}
\Phi_{\bm{p}}(\bm{x},t)=\frac{c}{n\cdot p}\int_0^{t_r}\dd\phi\Bigl[e\bm{A}(\phi)\cdot\bm{p}-\frac{1}{2}e^2\bm{A}^2(\phi)\Bigr],
\end{equation}
where the unit vector $\bm{n}=\bm{e}_2$ defines the pulse propagation direction, $t_r=t-\bm{n}\cdot\bm{x}/c$ is the retardation time, 
$n\cdot p=p^0-\bm{p}\cdot\bm{n}$, and $p\cdot x=cp^0t-\bm{p}\cdot\bm{x}$ with $p^0=\sqrt{\bm{p}^2+(\me c)^2}$.

In an approximation that takes into account only leading relativistic corrections of the order of $1/c$, this solution takes the form  \cite{PhysRevA.110.043112},
\begin{align}\label{sfa3}
\psi_{\bm{p}}(\bm{x},t)=\exp\bigl[&-\ii\me c^2t-\ii E_{\bm{p}}t+\ii\bm{p}\cdot\bm{x}+\ii\Phi_{\bm{p}}^{(0)}(t) \nonumber \\
&+\ii\pi_{\bm{p}}(t)\bm{n}\cdot\bm{x}+\frac{\ii}{\me c}\Phi_{\bm{p}}^{(0)}(t)\bm{n}\cdot\bm{p}\bigr], 
\end{align}
where
\begin{equation}\label{sfa4}
\Phi_{\bm{p}}^{(0)}(t)=\frac{1}{\me}\int_0^{t}\dd\phi\Bigl[e\bm{A}(\phi)\cdot\bm{p}-\frac{1}{2}e^2\bm{A}^2(\phi)\Bigr],
\end{equation}
and
\begin{equation}\label{sfa5}
\pi_{\bm{p}}(t)=-\frac{1}{\me c}\Bigl[e\bm{A}(t)\cdot\bm{p}-\frac{1}{2}e^2\bm{A}^2(t)\Bigr],
\end{equation}
with $E_{\bm{p}}=\bm{p}^2/2\me$. This means that the phase of the nonrelativistic Volkov state is modified by two terms present in the second line 
of Eq.~\eqref{sfa3}: $\pi_{\bm{p}}(t)\bm{n}\cdot\bm{x}$, which we call the retardation correction, and $\Phi_{\bm{p}}^{(0)}(t)\bm{n}\cdot\bm{p}/\me c$, 
called the recoil or the Nordsieck correction~\cite{PhysRev.93.785}. In order to investigate the influence of the retardation correction on 
multiphoton ionization, we consider the emission of photoelectrons along the laser pulse polarization vector, as in this case the 
recoil correction vanishes. The corresponding energy distributions are presented in Figs.~\ref{Interpol800x800XY2RPhi1Ca5IntE}(a) and (b). We clearly 
see that the retardation correction leads to the redshift of photoelectron energy distributions, but its effect is rather marginal for the considered laser 
pulse parameters. This agrees with our earlier investigations of ionization \cite{PhysRevA.92.043419,PhysRevA.94.013402} and laser-assisted radiative recombination \cite{PhysRevA.110.043112} processes, studied within the SFA formalism.

It appears that the recoil correction plays a much more significant role, as it follows from Figs.~\ref{Interpol800x800XY2RPhi1Ca5IntE}(c) and (d). 
To quantify its contribution let us assume that the laser field has a constant amplitude $A_0$; thus, enabling us to employ the quasi-energy 
picture. We further assume that, with the recoil correction neglected, the ionization distribution has a peak for a quasi-energy $E_n$ 
(which is nearly independent of the photoelectron emission angle $\varphi_{\bm{p}}$, as the retardation correction marginally contributes to 
the photoelectron energy distribution). Now, with the recoil correction accounted for, this quasi-energy is shifted. As one can derive for a plane wave
field using Eqs.~\eqref{sfa3} and~\eqref{sfa4}, the peak appears for the angle-dependent quasi-energy,
\begin{equation}\label{sfa6}
E_n(\varphi_{\bm{p}})=E_n-\Omega_n\sin\varphi_{\bm{p}},
\end{equation}
where
\begin{equation}\label{sfa7}
\Omega_n=U_p\frac{\sqrt{2\me E_n}}{\me c}=U_p\sqrt{\frac{2E_n}{\me c^2}},
\end{equation}
and $U_p=e^2A_0^2/4\me$ is the ponderomotive energy for the plane wave of a constant intensity. Note that this result is valid up to
the leading nondipole order in $1/c$.

The energy $\Omega_n$ can be used for quantifying the importance of nondipole effects in ionization. Namely, these effects are moderately important 
(and so the laser field is considered to be moderately intense) if $\Omega_n\leqslant\omega/2$, as in this case the contributions coming from the 
$n$-th peak to the angle-integrated energy distribution $\tilde{P}_m(|\bm{p}|,T_\mathrm{f})$ are clearly separated from the contributions for the adjacent 
$n\pm 1$ peaks. In other words, in the energy distribution we clearly distinguish the multiphoton structures, independently of the fact that these structures 
can be significantly modified by nondipole effects. On the other hand, the nondipole effects become important for strong laser fields, for which 
$\Omega_n >\omega/2$. In this case, the nondipole corrections from the $n$-th peak overlap with those emerging from the adjacent multiphoton peaks. 
Note, however, that this classification of the field strength applies only to long pulses with slowly changing in time envelopes, as under these 
circumstances in the dipole approximation we observe well separated multiphoton peaks in photoelectron energy distributions. Let us also remark that the 
above classification depends on the photoelectron energy. Specifically, for highly energetic photoelectrons this is a rather rough 
estimation, as it accounts only for the leading relativistic corrections. We stress that the multiphoton structures could still be observed in such case, 
with however a very low signal. In contrast, for weakly intense laser fields we have 
$\Omega_n\ll\omega$. In this case, the relativistic effects are tiny, although for the present day computational techniques they are clearly
observed in numerical analysis (see, e.g., the extended analysis presented in Refs.~\cite{PhysRevA.109.013107,Vembe_2024}, which is discussed below).

Coming back now to the analysis presented in Sec.~\ref{sec:ModeratePulse}, indeed, we observe there the well separated multiphoton peaks in the energy 
distribution, cf. Figs.~\ref{Interpol800x800XY2RPhi1Ca5IntE0}(a) and (b). However, the nondipole corrections significantly modify their shapes, 
especially for the middle range of energies, for which we observe the double-hump structures. Next, let us choose energies corresponding 
to particular multiphoton peaks, as indicated by the white lines in Fig.~\ref{Interpolation800x800XY2RPhi1Cda5K}(b). These two lines correspond 
to photoelectron energies $27.75E_0$ and $31.75E_0$, whereas in Fig.~\ref{Interpolation800x800XY2RPhi1Ca5K}(b) we draw the corresponding two red 
lines by applying Eq.~\eqref{sfa6} with $E_n$ equal to each of these values. As we see, the red lines very well reproduce the $\varphi_{\bm{p}}$-dependent 
peaks in the photoelectron momentum distribution $\tilde{P}_m(|\bm{p}|,\varphi_{\bm{p}},T_\mathrm{f})$. In this case, i.e., for a moderately
intense laser field, the SFA formalism well describes the wavy pattern observed in the energy-angular distribution of photoelectrons.

\begin{figure}[t]
\includegraphics[width=7cm]{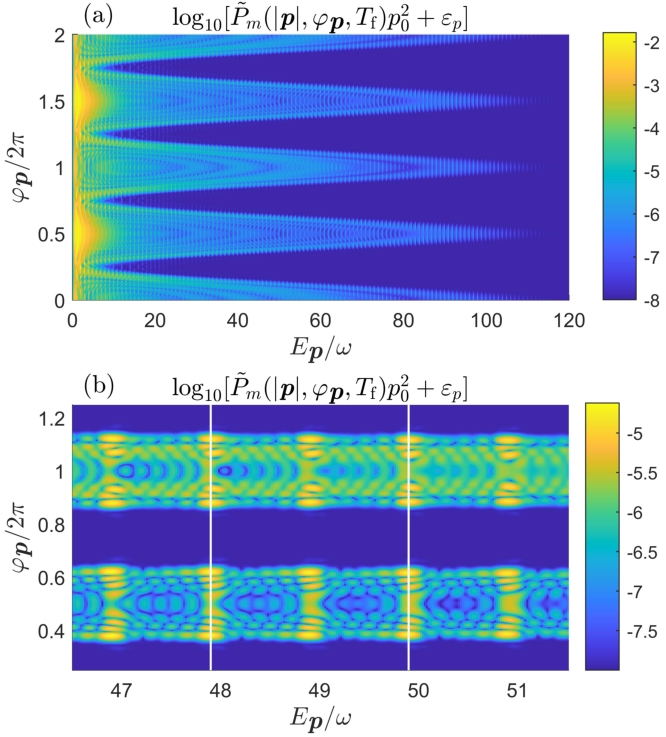}
\caption{Photoelectron momentum distributions in the polar coordinates calculated within the dipole approximation. The parameters of numerical 
analysis are defined in Sec.~\ref{sec:StrongPulse}. For visual purposes, in panel (a) we present two copies of the distribution, i.e., for
$0\leqslant\varphi_{\bm{p}}\leqslant 4\pi$. A closeup portion of this distribution is shown in panel (b). For clarity, the values smaller than 
$\varepsilon_p=10^{-8}$ have been removed from both panels. The meaning of two vertical lines is discussed in Sec.~\ref{sec:StrongPulse}.
}
\label{Interpolation1100x800XY2RPhi1Cda10K}
\end{figure}
\begin{figure}[t]
\includegraphics[width=7cm]{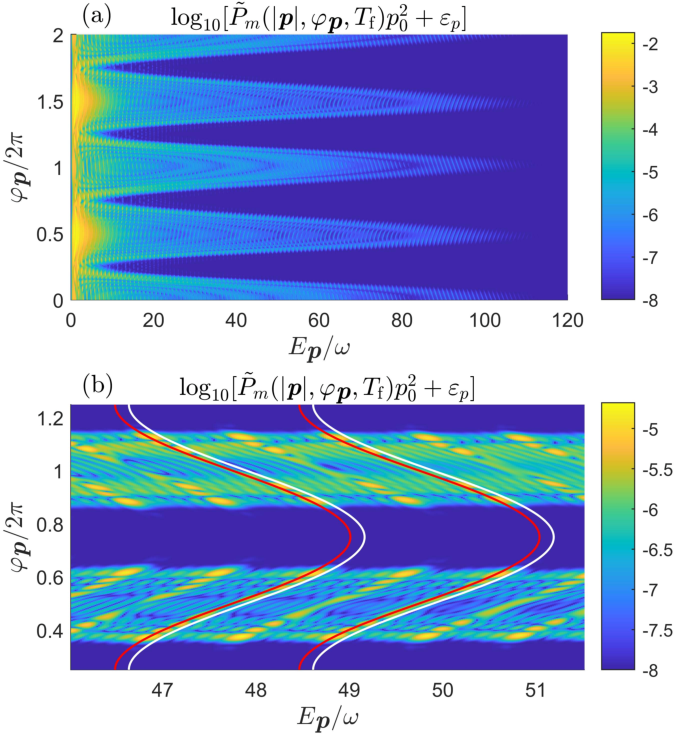}
\caption{The same as in Fig.~\ref{Interpolation1100x800XY2RPhi1Cda10K}, but for the laser pulse propagating along the $x_2$-axis.
The plotted wavy lines are discussed in Sec.~\ref{sec:StrongPulse}.
}
\label{Interpolation1100x800XY2RPhi1Ca10K}
\end{figure}

Finally, let us comment on the results of Refs.~\cite{PhysRevA.109.013107,Vembe_2024}. They are calculated for the laser pulse with the sin$^2$ 
envelope, the central frequency $\omega=50E_0$, and the maximum electric field strength $\mathcal{E}_\mathrm{max}=600\mathcal{E}_0$. 
This means that the temporal ponderomotive energy of the pulse does not exceed the value $U_p=e^2\mathcal{E}^2_\mathrm{max}/(4\omega^2)=36E_0$. 
According to our analysis, such a laser pulse can be considered as a moderately intense or even a weak one, as $\Omega_n/\omega\approx 0.13$ 
for the photoelectron energy $300E_0$ considered in~\cite{PhysRevA.109.013107,Vembe_2024}. For this reason, the integrated over the angles 
energy distribution of photoelectrons consists there of multiphoton peaks, lacking however the double-hump structure observed in our analysis. Independently of this, our investigations 
fully agree with the conclusions drawn in Refs.~\cite{PhysRevA.109.013107,Vembe_2024}. Namely, that for the energy distributions and in the nonrelativistic 
case the laser pulse propagation corrections lead to the redshift of energy peaks. The aim of the next section is to investigate to what extent these conclusions remain valid for strong laser pulses.

\section{Strong laser pulses}
\label{sec:StrongPulse}

We consider a ten-cycle laser pulse ($N_\mathrm{osc}=10$), with eight cycles comprising the flat-top portion of the envelope. The pulse central 
frequency is the same as above, $\omega=2E_0$, whereas its amplitude $A_0$ is changed such that $|e|A_0=10p_0$. This means that the ponderomotive
energy within the flat top equals $U_p=e^2A_0^2/4\me=25E_0$. The convergence of numerical calculations is achieved for the following parameters: 
$x_{10}=1100a_0$, $x_{20}=800a_0$, $K_1=14$, $K_2=13$, and $\delta t=0.01t_0$. For the mask function $M(x_1,x_2)$ we choose $r_1=1050a_0$ and 
$r_2=1099a_0$. For the chosen parameters, the laser pulse lasts for $T_\mathrm{p}=2\pi N_\mathrm{osc}/\omega=31.4t_0$. Therefore, the final time of the numerical 
calculation is chosen to be $T_\mathrm{f}=43.54t_0$. Hence, the probability that the electron escapes the integration region is smaller than 
$1.2\times 10^{-8}$, whereas the probability for the electron to stay in the ground state equals $0.07$. For the mask function $M_0(x)$ we select 
the same values as above, $R_1=30a_0$ and $R_2=90a_0$. Note that we have to decrease the number of laser pulse cycles as compared to the moderately 
intense pulse considered in Sec.~\ref{sec:ModeratePulse}, in order to keep the same order of magnitude for the electron wave function leakage 
from the integration region. The current laser pulse, according to the classification described in Sec.~\ref{sec:SFAanalysis}, 
is considered to be strong for electron energies $E_{\bm{p}}>\me c^2\omega^2/8U_p^2\approx 15E_0$.

In Figs.~\ref{Interpolation1100x800XY2RPhi1Cda10K} and \ref{Interpolation1100x800XY2RPhi1Ca10K}, we present the energy-angular distributions 
$\tilde{P}_m(|\bm{p}|,\varphi_{\bm{p}},T_\mathrm{f})$ calculated in the dipole approximation and for the propagating in space pulse, respectively. 
Qualitatively, we observe the same pattern as for the moderately intense pulse, except that now the inclination of stripes relative to the horizontal 
direction is smaller. The interpretation of this fact can be based on the SFA analysis presented in the previous section. To this end, we choose 
two energies $E_1^{(d)}=95.8E_0$ and $E_2^{(d)}=99.8E_0$ represented in Fig.~\ref{Interpolation1100x800XY2RPhi1Cda10K}(b) by two white lines. 
Approximately for these energies we observe the multiphoton peaks in the energy distribution $\tilde{P}_m(|\bm{p}|,T_\mathrm{f})$ calculated in 
the dipole approximation. The recoil correction transforms these straight lines, according to Eq.~\eqref{sfa6}, into the sinusoidal curves.
The latter are represented by white lines in Fig.~\ref{Interpolation1100x800XY2RPhi1Ca10K}(b). The amplitude of these sinusoidal curves is 
larger than $\omega$, hence, according to the classification introduced in Sec.~\ref{sec:SFAanalysis}, this pulse is considered to be strong. 
However, a closer look at this figure shows that the white lines do not follow sufficiently well the 'tilted stripes'. The fit is corrected if 
we choose for the propagating laser pulse the following values: $E_1^{(p)}=95.5E_0$ and $E_2^{(p)}=99.5E_0$. The corresponding sinusoidal curves 
are represented in Fig.~\ref{Interpolation1100x800XY2RPhi1Ca10K}(b) by red lines, which now much better suit to the stripe pattern. The reason 
for such a redshift is the retardation correction discussed in Sec.~\ref{sec:SFAanalysis}.

\begin{figure}[t]
\includegraphics[width=7cm]{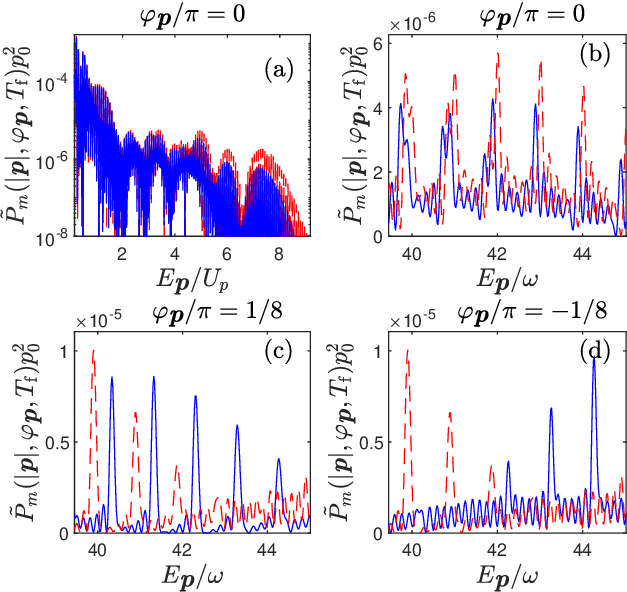}
\caption{The same as in Fig.~\ref{Interpol800x800XY2RPhi1Ca5IntE}, but for the strong laser pulse characterized by the parameters 
specified in Sec.~\ref{sec:StrongPulse}.
}
\label{Interpol1100x800XY2RPhi1Ca10IntE}
\end{figure}
\begin{figure}[t]
\includegraphics[width=7cm]{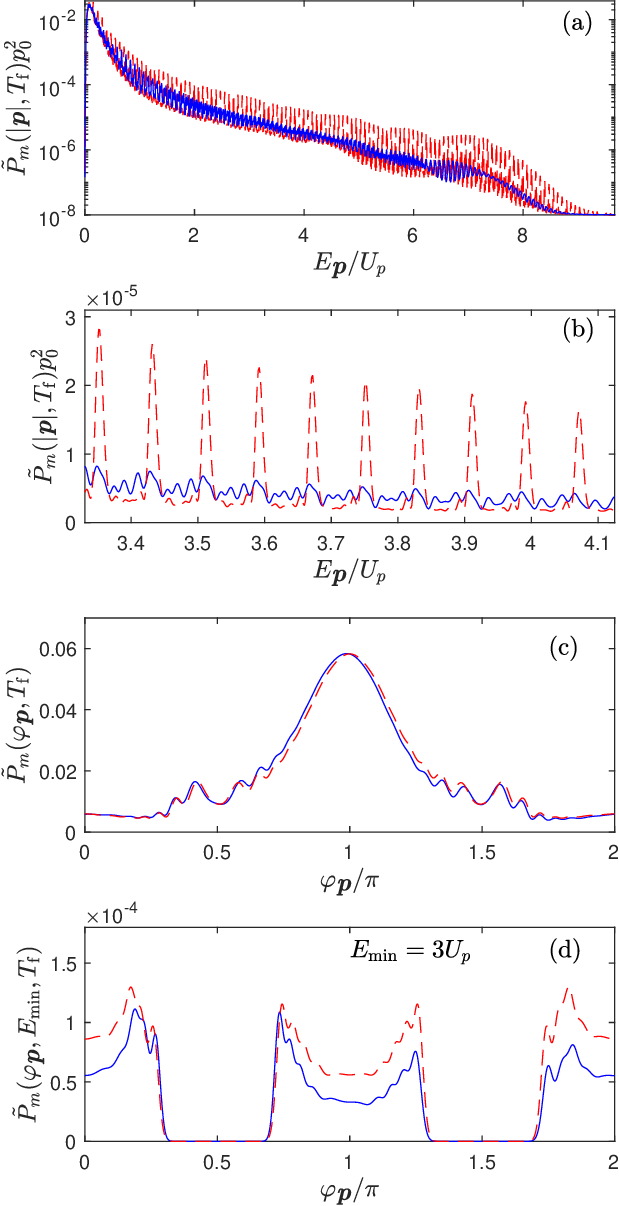}
\caption{The same as in Fig.~\ref{Interpol800x800XY2RPhi1Ca5IntE0}, but for the strong laser pulse specified in Sec.~\ref{sec:StrongPulse}.
}
\label{Interpol1100x800XY2RPhi1Ca10IntE0}
\end{figure}

As mentioned in Sec.~\ref{sec:SFAanalysis}, the retardation effect is most clearly manifested for photoelectrons ionized in the direction 
of the laser pulse polarization. This is presented in Figs.~\ref{Interpol1100x800XY2RPhi1Ca10IntE}(a) and (b) as well, where the photoelectron spectra
calculated for a propagating laser pulse (solid lines) are plotted against the spectra obtained within the dipole approximation (dashed lines). 
As for the moderately intense pulses, in both cases we observe the formation of the rescattering plateau ranging from $2U_p$ up to $8U_p$. 
Moreover, the inspection of Fig.~\ref{Interpol1100x800XY2RPhi1Ca10IntE}(b) shows that the retardation correction leads to a small redshift of 
the spectrum, as discussed in Refs.~\cite{PhysRevA.109.013107,Vembe_2024}. However, much more pronounced red- and blueshift occur for emission 
directions different from the polarization one, as presented in Figs.~\ref{Interpol1100x800XY2RPhi1Ca10IntE}(c) 
and (d), and which is the effect related to a more significant recoil effect.

Marginal distributions are demonstrated in Fig.~\ref{Interpol1100x800XY2RPhi1Ca10IntE0}. As for the moderately intense pulse, we observe the formation 
of rescattering plateau, Fig.~\ref{Interpol1100x800XY2RPhi1Ca10IntE0}(a), with the cutoff for photoelectron energies around $8U_p$. However, details 
presented in Fig.~\ref{Interpol1100x800XY2RPhi1Ca10IntE0}(b) prove that the multiphoton structure survives only in the dipole approximation, which is due to 
the overlap of the 'tilted' multiphoton stripes shown in Fig.~\ref{Interpolation1100x800XY2RPhi1Ca10K}. Moreover, the high-energy portion of the 
spectrum is now smaller in magnitude and the rescattering plateau for increasingly stronger pulses gradually disappears. This follows from the fact that the stronger pulses 
push the electron wave packet away from the parent ion and so the probability of electron rescattering decreases. Hence, the ionization dynamics 
is not any longer dominated by the in-pulse rescattering but rather by the post-pulse spreading of the electron wave 
packet, as already discussed in Refs.~\cite{PhysRevA.107.053112,Suster:24}.

For the angular distribution of photoelectrons, shown in Fig.~\ref{Interpol1100x800XY2RPhi1Ca10IntE0}(c), we observe the same pattern as for the 
moderately intense pulse. Namely, the probability of emission in the direction of the pulse propagation is enhanced by nondipole corrections, 
as oppose to the reversed direction. However, in the case of high-energy photoelectrons, the dipole approximation always overestimates the 
angular momentum distributions [Fig.~\ref{Interpol1100x800XY2RPhi1Ca10IntE0}(d)]. Again, this indicates that the in-pulse rescattering processes, 
resulting in high-energy photoelectrons, are less probable for intense propagating laser fields. 

\section{Conclusions}
\label{sec:conclusion}

In this paper, we have studied nondipole effects in multiphoton ionization. For this purpose, we have solved numerically a two-dimensional
Schr\"odinger equation with a model atom interacting with a relatively long flat-top pulse. This has been done for moderately intense
and for intense laser pulses. In both cases, the results accounting for pulse propagation have been confronted with the analogous results 
obtained within the dipole approximation. By presenting the photoelectron energy-angular distributions in polar coordinates, 
we were able to address the directionality dependence of the spectra for the former. For the given laser field parameters, our numerical results
turned out to be in agreement with an analytical estimate based on the leading order relativistic expansion of the electron Volkov state.
We confirmed in this way that the electron recoil off the laser pulse is a deciding factor that characterizes the directionality dependence
of the energy spectra of photoelectrons, whereas the retardation correction leads to its tiny redshift. While studying the marginal distributions
of photoelectrons we have also observed a double-hump structure of multiphoton peaks, that occurs up to the mid-energy region in the spectrum.
Also, with increasing the laser field intensity, we have seen disappearance of multiphoton peaks as well as diminishing of the plateau region.
Such features indicate that the in-pulse dynamics of the electron resulting in its rescattering becomes less probable for intense propagating
laser pulses. A more detailed analysis of the electron dynamics in this case can be found in Refs.~\cite{PhysRevA.104.L021102,PhysRevA.107.053112,Suster:24}.

According to our analysis, for the current parameters, the leading order relativistic corrections accurately describe the physics governed 
by the Hamiltonian~\eqref{ps4}. It is worth noting, however, a series of papers with a focus on multiphoton ionization by x-ray laser 
pulses~\cite{PhysRevA.109.013107,Vembe_2024,PhysRevA.99.053410,PhysRevA.101.063416,PhysRevA.106.013104,PhysRevLett.121.253202}. In these papers, it was proven that the higher order 
relativistic corrections cannot be neglected in describing the ionization dynamics for their given parameters. While similar analysis  
is beyond a scope of our paper, it can be performed in the future.

\section*{Acknowledgements}

We thank Morten F\"orre, S\o lve Selst\o, and Johanne E. Vembe for illuminating discussions of nondipole effects in ionization. 
We also acknowledge hospitality of the Oslo Metropolitan University, Norway, during the Workshop: Atoms in Strong Laser Fields. This work was supported 
by the National Science Centre (Poland) under Grant No. 2018/30/Q/ST2/00236.

\bibliography{st1biblio}

\end{document}